\documentclass[12pt]{article}

\usepackage[a4paper,left=1.5cm,right=1.5cm,top=2.5cm,bottom=2.5cm]{geometry}
\usepackage{graphicx}
\usepackage{booktabs}
\usepackage{listings}
\usepackage{hyperref}

\usepackage[comma]{natbib}
\usepackage{breakcites}
\usepackage{placeins}
\usepackage{amsmath}
\usepackage{verbatim}
\usepackage{amssymb}
\usepackage{multicol}
\usepackage{color}
\usepackage[english]{babel}
\usepackage{siunitx}
\usepackage{graphicx}
\usepackage{wrapfig}
\usepackage{comment}
\usepackage{amsmath,empheq}
\usepackage{amssymb}
\usepackage{wasysym}
\usepackage{multirow}
\usepackage{lscape}
\usepackage[margin=1cm]{caption}
\usepackage{subfigure}
\usepackage[subfigure]{tocloft}
\usepackage[english=british]{csquotes} 

\usepackage{pdflscape}
\usepackage{blindtext}

\makeatletter
\ifcsname phantomsection\endcsname
    \newcommand*{\qrr@gobblenexttocentry}[5]{}
\else
    \newcommand*{\qrr@gobblenexttocentry}[4]{}
\fi
\newcommand*{\addsubsection}{
    \addtocontents{toc}{\protect\qrr@gobblenexttocentry}
    \subsection}
\makeatother

\title{Transition from geostrophic flows to inertia-gravity waves in the spectrum of a differentially heated rotating annulus experiment \footnote{This work has been submitted to \textit{Journal of the Atmospheric Sciences (JAS)}. Copyright in this work may be transferred without further notice. }.}
\author{Costanza Rodda, Uwe Harlander}
\date{\small{\textit{Department of Aerodynamics and Fluid Mechanics, Brandenburg University of Technology Cottbus-Senftenberg} }}
\begin{document}
\maketitle{}

\abstract{Inertia-gravity waves (IGWs) play an essential role in the terrestrial atmospheric dynamics as they can lead to energy and momentum flux when propagating upwards. An open question is to which extent IGWs contribute to the total energy and to flattening of the energy spectrum observed at the mesoscale.
In this work, we present an experimental investigation of the energy distribution between the large-scale balanced flow and the small-scale imbalanced flow. Weakly nonlinear IGWs emitted from baroclinic jets are observed in the differentially heated rotating annulus experiment. Similar to the atmospheric spectra, the experimental kinetic energy spectra reveal the typical subdivision into two distinct regimes with slopes $k^{-3}$ for the large scales and $k^{-5/3}$ for the small scales. By separating the spectra into the vortex and the wave component, it emerges that at the large-scale end of the mesoscale the gravity waves observed in the experiment cause a flattening of the spectra and provide most of the energy. At smaller scales, our data analysis suggests a transition towards a turbulent regime with a forward energy cascade up to where dissipation by diffusive processes occurs.}

%

 \section{Introduction}
%
Atmospheric motions at the mid-latitudes can be divided into synoptic-scale balanced motions, which develop from the baroclinic instability of the westerly flow, and mesoscale unbalanced motions, among which there are inertia-gravity waves (IGWs). It is now established that the balanced motions cannot exist without emitting IGWs in analogy to an elastic pendulum that cannot swing without fast axial oscillations \citep{vanneste2013balance}. This IGW generation process, observed in regions of the jet-stream \citep{o1995generation}, is called `spontaneous emission' and it is considered one of the relevant sources in the atmosphere together with IGW generation by orography and convection. 

The synoptic- and mesoscale motions have distinct characteristics, and their time scales are usually well separated. The different dynamic regimes are reflected in the structure of the atmospheric kinetic energy spectra as shown by \citet{nastrom1985climatology}, who provided the first comprehensive spectra of the zonal and meridional wind components and temperature measured near the tropopause level by commercial aircraft. The spectra exhibit two distinct power-law dependencies upon wavenumber in the form $\mathcal{P}\propto k^{p_{k}}$, where the typical measured values are $p_{k}=-3$ in the synoptic-scale range (for wavelengths between $500$ km and $3000$ km), and $p_{k}=-5/3$ in the mesoscale range (for wavelengths smaller than $500$ km). 
The steep slope corresponding to the $-3$ power-law exhibited by the quasi-geostrophic balanced flow, is consistent with Charney's theory of geostrophic turbulence. Therefore, the hypothesis for this range of the spectrum is a downscale potential enstrophy (QG) cascade.
For wavelengths smaller than $500$ km, the  spectral slope flattens and approaches $-5/3$; this suggests that at the mesoscale there is a different dynamical regime and some other phenomena are responsible for this part of the spectrum. Many investigations over the past years have focused on the mesoscale energy sub-range trying to explain the phenomena involved and whether the energy is forward or inverse cascading (see, for example, \citet{lindborg2007horizontal}, \citet{waite2009mesoscale}, \citet{callies2014transition} and references therein). Observational evidence \citep{cho2001horizontal} indicated a downward energy cascade and put an end to the debate about the direction of energy at the mesoscales \citep{gage1979evidence, lilly1983stratified}. However, the energy source at these smaller scales remains a subject of debate. Although there is not a consensus on the nature of the mesoscale part of the spectra, the impact of spontaneously generated gravity waves on the mid-latitude dynamics is well established, and it is, therefore, natural to include them among other possible sources \citep{dewan1979stratospheric,callies2014transition,vzagar2017energy}. Together with gravity waves, the other most invoked mechanisms are strongly stratified turbulence \citep{lindborg2006energy} and quasi-geostrophic turbulence \citep{tung2003k}.\\

A method that has been used to test the contribution of gravity waves to the energy consists in the separation of the kinetic energy spectrum into divergent and rotational parts \citep{buhler2014wave, lindborg2015helmholtz}. If the divergent part dominates and its ratio to the rotational component is larger than one, IGWs are the leading energy's contributors. By applying this separation method to subsets of the MOZAIC (Ozone and Water Vapor by Airbus In-Service Aircraft) dataset, \cite{callies2014transition} found a dominance of the divergence spectrum at the mesoscale while \cite{lindborg2015helmholtz} identified the rotational part as predominant and therefore concluded that stratified turbulence, rather than IGWs, is the prevailing mechanism. One possible reason behind these conflicting results is that although the energy spectra display global characteristics, the dynamics underlying the mesoscale energy spectrum may vary between different regions of the atmosphere \citep{bierdel2016accuracy}.\\

In this paper, we propose a laboratory investigation of inertia-gravity waves emitted from baroclinic jets. One of the main advantages of using a laboratory experiment is that some gravity wave sources like orography or convection can be easily excluded, contrarily to measured atmospheric data that contain all sources of IGWs. The differentially heated rotating annulus experiment, a well-established experiment to study an analogue dynamics of the atmosphere \citep{read2014general}, is used to generate baroclinic waves from which inertia-gravity waves are emitted \citep{hien2018spontaneous, rodda2018baroclinic, Rodda2019}. The properties of the experimentally observed gravity waves are studied in detail together with the energy distribution among the balanced and imbalanced flow. The outcome of our investigation is a simplified picture of the atmospheric multiple-scale flow for which the relevant processes and spectral properties are more accessible than for atmospheric field data. 

The paper is structured in the following way. In section \ref{sec:data}, the experimental set-up of the differentially heated rotating annulus and the measurement techniques are illustrated. Section \ref{sec:IGWs} focuses on the gravity wave properties observed experimentally. The typical location of IGWs with respect to the jet front, their propagation velocity, horizontal scale and frequencies are examined in detail. The energy distribution among the scales is investigated in section \ref{sec:spectra}. To evaluate the contribution of different phenomena, we separate the kinetic and total energy spectra into geostrophic and gravity wave components. At the end of the section, some considerations on the validity of the weakly nonlinear regime assumption used for the flow separation are discussed. In section \ref{sec:conclusions}, a summary of our main findings and conclusions are given.
\section{Experimental set-up and data acquisition}
\label{sec:data}
The experiment used in this work is an atmospheric-like differentially heated rotating annulus for which inertia-gravity waves emitted from the baroclinic jet have been observed both numerically \citep{borchert2014gravity} and experimentally \citep{Rodda2019}. A plexiglass tank subdivided into three concentric cylinders, with inner radius $a=35$ cm and outer radius $b=70$ cm, is mounted on a turntable and filled with de-ionised water. The water in the inner cylinder is kept to a constant low temperature by a system of pumps connected to a thermostat. Similarly, the water in the outer cylinder is kept to a constant higher temperature. The middle gap is our investigation region, and it is filled with water to a total depth of $H=6$ cm. The water in the gap, initially at room temperature, experiences a radial temperature difference imposed by the two insulated walls and consequently convective rolls covering the entire fluid depth develop. For high enough rotation rates, the flow becomes baroclinically unstable and forms baroclinic waves. A more detailed description of the experiment and the flow regimes can be found in \citet{Rodda2019}. %

The set of experiments presented in this paper is run with the lateral temperature difference set to $\Delta T=4.8$ K and the rotation rate varying from $\Omega=0.5$ rpm to $1$ rpm. 
The measurement technique used to investigate the velocity of the flow in the mid-gap is the non-intrusive two-dimensional Particle Image Velocimetry (PIV). The seeding particles are hollow glass spheres from DANTEC with mean particle size $10$ $\mu$m and density $1.1$ g cm$^{-3}$. A diode-pumped steady laser (LINOS nano, max power $100$ mW, wavelength $532$ nm) is fixed to the tank and co-rotates with it. The laser beam goes through a cylindrical-lens optical system that spreads it into a 2D light plane  of thickness $\approx 1$ mm, which is small enough to be approximated to a two-dimensional plane. The area illuminated by the laser (covering approximately 1/6 of the annulus), is recorded by a co-rotating GoPro hero 4 black camera (video resolution is set to 1080p, with 48 fps). The camera is updated with a lens (The Imaging Source TCL 1216) that allows for control on the focus and the aperture. The recorded videos have a total duration between 45 and 50 minutes for each experimental run. 

The frames recorded by the camera are successively processed with sophisticated image-processing techniques that calculate the  displacement of the particle pattern illuminated by the laser between two images and output the fluid velocities. For this data processing, we use the free Matlab toolbox UVmat, developed at LEGI in Grenoble (downloadable at \url{http://servforge. legi.grenoble-inp.fr/projects/soft-uvmat}). The complete documentation about the use of the software can be found in \citet{SommeriaUVmat}.
Finally, the obtained velocities are organised on a regular cartesian grid with a spatial resolution of $0.2$ cm in both $x-y$ axis. Furthermore, a time average is applied to reduce the high-frequency noise and, therefore, the final temporal resolution is 0.5 s.
%
\section{Inertia-gravity waves}
\label{sec:IGWs}
%
\begin{figure}[t]
\centering
\includegraphics[width=0.5\textwidth]{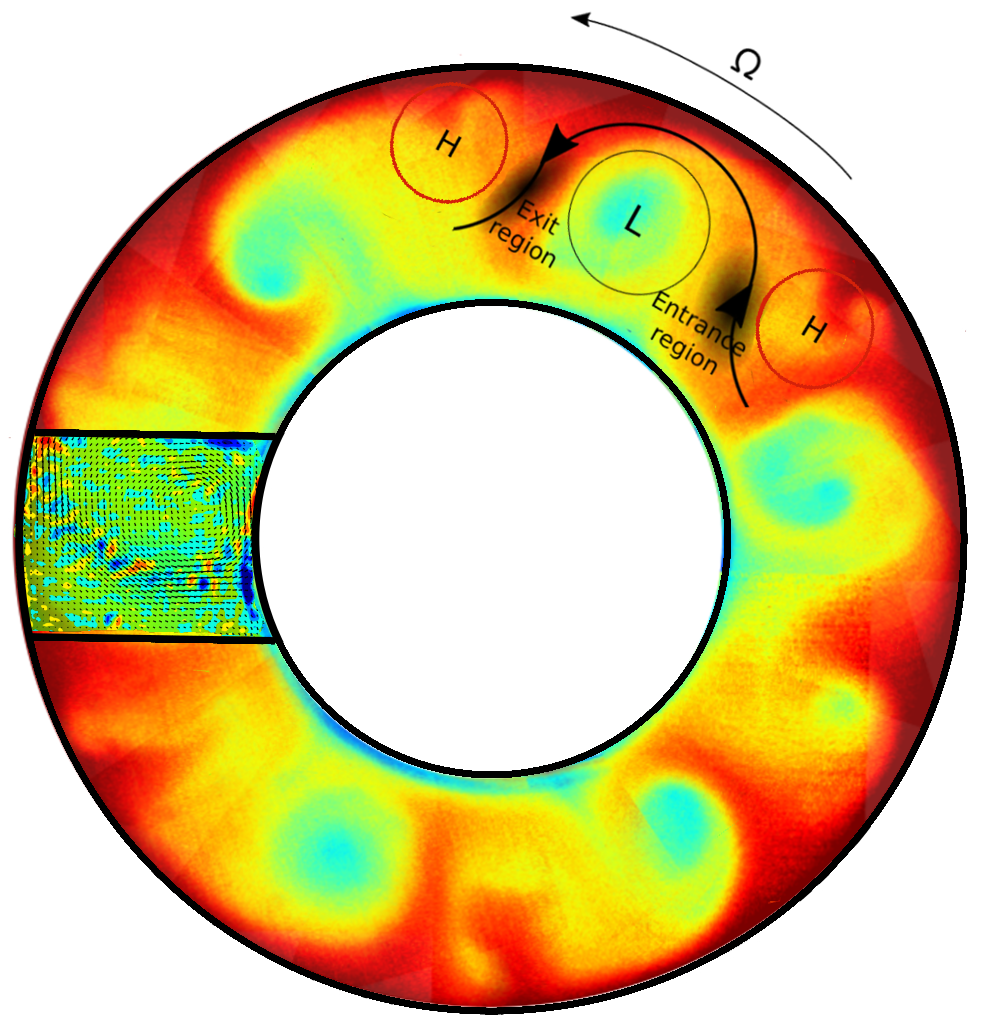}
 \caption[Combined plot of the surface temperature and the velocity field]{Combined plot of the surface temperature reconstructed from measurements done with the infrared camera, and the velocity field (black arrows). The sketch illustrates the position of the entrance and exit region of the jet. The baroclinic wave is in a regular regime with azimuthal wavenumber $m=6$. The colours in the small inlet show the horizontal divergence measured by PIV. Small-scale wavelike features are recognisable in the horizontal divergence field; they appear along the baroclinic jet and propagate with it.}
\label{fig:DivThermo}
\end{figure}
Inertia-gravity waves, due to their small spatial and short temporal scales, are challenging to observe directly in the velocity field, which is dominated by the large-scale flow. However, since in good approximation, such a balanced flow can be considered divergence-free, the imbalanced part of the flow can be identified as wave pattern in the horizontal divergence. For this reason, many studies employ the horizontal divergence as a dynamical indicator of IGWs \citep{o1995generation,plougonven2003inertia, wu2004study, dornbrack20122009, khaykin2015seasonal}.

A snapshot of the horizontal divergence calculated from the horizontal velocity components (measured at a fluid height $z=5$ cm for an experimental run with $\Omega=0.5$ rpm) is plotted in figure \ref{fig:DivThermo} in the inset on the left. The plot is superimposed to the surface temperature map measured with an IR camera to help to visualise the small-scale features position at the exit region of the large-scale baroclinic wave (see sketch representing the jet exit and entrance regions in figure \ref{fig:DivThermo}). The largest amplitude in the horizontal divergence signal is concentrated along the jet front and exhibits wavelike structures with the wave crests roughly perpendicular to the flow. They appear irregularly, propagate within the jet stream, are mostly advected by it, and then dissipate; we refer to this irregular behaviour as intermittency. The small waves present similar characteristics also for the other experimental runs at different rotation rates varying in the range $0.5$ rpm $<\Omega<1$ rpm. In all experiments, the regimes explored are characterised by baroclinic waves with a dominant spatial structure and azimuthal wavenumber $m=5,6,7$ depending on the rotation rate. However, in some cases, we have steady wave regimes while in others, the waves show amplitude vacillations. The situation for full geostrophic turbulence is not included in the present study.

A rough estimate of the zonal and meridional horizontal wavelength from figure \ref{fig:DivThermo} gives $\lambda_{x}\approx 5$ cm and $\lambda_{y} \approx 4.5$ cm, in the local cartesian system of reference, where $y$ points in the radial direction. If we compare these wavelengths with the typical length scale of the baroclinic vortices, which is given by the Rossby deformation radius $L_{D}=NH/f\approx 12$ cm, it emerges that IGWs are approximately three times smaller than the core region of the baroclinic vortices. These findings are in agreement with \citet{kafiabad2018spontaneous}, who evinced that the spontaneous breakdown of the balanced flow and energy transfer to IGWs occurs when the latter is in the order of magnitude of the larger-scale balanced flow.\\

The small-scale waves propagating along with the jet qualitatively resemble the atmospheric inertia-gravity waves observed along jets and fronts. In particular, observations but also numerical models indicate the jet exit region and also, although less frequently, the entrance region as favoured locations for large-amplitude IGWs (see, e.g. \citet{plougonven2014internal} and references therein, \citet{dornbrack2018gravity}, and \citet{von2019interior}). 

Another important characteristic of the jet's exit region is the large deformation of the flow due to its deceleration. It has been observed by \citet{plougonven2005gravity} that regions with intense horizontal deformation and vertical shear can be crucial in the gravity wave propagation giving rise to a phenomenon called wave capture \citep{buhler2005wave}. When wave capture occurs, the horizontal wave vectors tend to align with the contraction axis of the flow, whereas the tilt of wave vectors tends to converge to a value given by the ratio of vertical shear and deformation \citep{wang2009generation}. It can be noticed in figure \ref{fig:DivThermo} that the wave trains have, indeed, horizontal wave vectors that tend to orient along the contraction axis, which is aligned with the jet, and occur in regions where the deformation is the largest. This investigation reveals that the IGWs emitted from the baroclinic flow in our experiment likely undergo wave capture during their propagation through the flow \citep{Rodda-thesis}.

\subsection{IGWs dispersion relation}
\label{IGW-dispersion}
\begin{figure}[t]
  \centering
\includegraphics[width=0.8\linewidth]{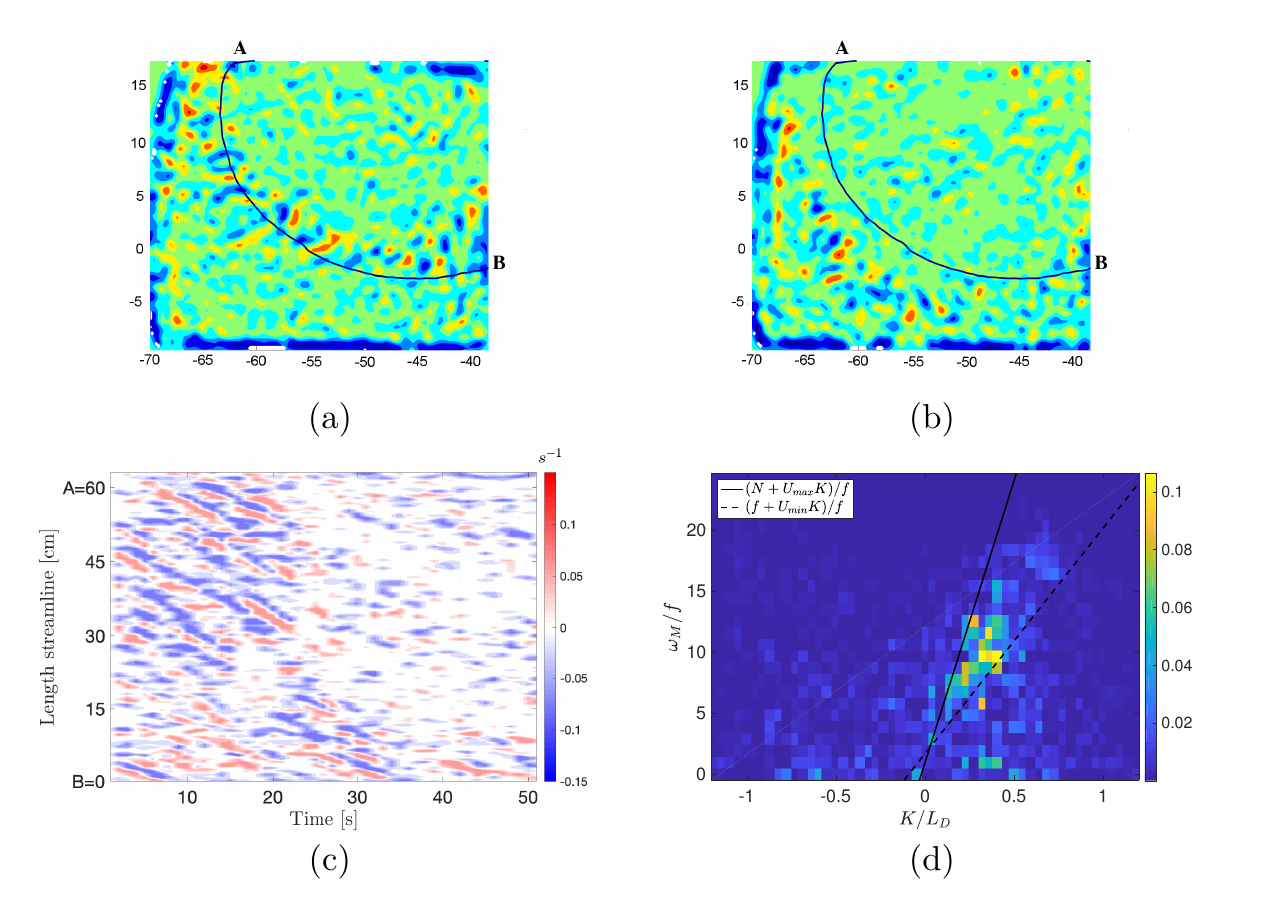}
 \caption[Frequencies along a fixed streamline]{Analysis of the frequencies and wavenumbers. (a) and (b) display the plot of the horizontal divergence and the streamline at $t_{0}=0$ and $t_{1}=40$ s. (c) Hovm\"{o}ller plot of the horizontal divergence for 1 s $\le t \le 50$ s, and (d) 2D Fourier transform of (c). The measured frequencies and wavenumbers are normalised by the Coriolis frequency $f=0.13$ rad s$^{-1}$ and the Rossby deformation radius $L_{D}=12$ cm. The black solid and dashed lines depict the the upper and lower limits for IGWs, which are $f$ and $N=0.22$ rad s$^{-1}$ doppler shifted using the maximum velocity $U_{\text{max}}=0.5$ cm s$^{-1}$ and minimum velocity $U_{\text{min}}=0.2$ cm s$^{-1}$ measured along the jet and multiplied by $-15$ cm$^{-1}$ $<K<15$ cm$^{-1}$, respectively.}
   \label{fig:Div1stream}
\end{figure}

To further investigate the properties of the observed wave train structures, we perform an analysis in Fourier space. The aim is to identify the wave frequencies and see if they fulfil the IGWs dispersion relation
\begin{equation}
 \omega_{i}^2= \frac{N^{2}K_{H}^{2}+f^{2}n^{2}}{K_{H}^2+n^2}, \label{eq:IGWdisp}
\end{equation}
where $\omega_{i}$ is the intrinsic frequency, $N$ is the buoyancy frequency, $f=2 \Omega$ the Coriolis frequency, $K_{H}=\sqrt{k^2+l^2}$ the horizontal wavenumber, and $n$ the vertical wavenumber. Frequencies and wavenumbers of the measured waves can be obtained by calculating the space-time data Fourier transform. The two-dimensional data for such Fourier analysis consist of the horizontal divergence sampled along the jet over time. This choice is the most convenient considering that, as we have previously shown, the small-scale waves are aligned with the jet and propagate within it. The data along the jet are calculated by interpolating the horizontal divergence along a streamline frozen in time, identifying periods where the jet front is passing over the chosen streamline. The function defining streamlines is the streamfuncion $\Psi$ that is related to the horizontal velocity components as $u= \partial_{y} \Psi$ and $v= - \partial_{x} \Psi$.
Once the streamfunction is known, the streamlines are straightforwardly identified by a unique constant. One streamline fixed in space, identifying the baroclinic front at $t=t_{0}$, and the horizontal divergence data are plotted in figure \ref{fig:Div1stream}(a) and (b) for $t_{0}$ and $t_{1}=t_{0}+40$ s respectively.
The horizontal divergence is then interpolated along this curve for the time interval considered and successively displayed in a time-space Hovm\"oller plot (see figure \ref{fig:Div1stream}(c)). The orientation of the divergence signal in figure \ref{fig:Div1stream}(c) shows no alterations in the slope, meaning that the waves propagate with the same angle to the jet and with a constant phase speed. Moreover, the signal fades in time, showing that the strongest waves are embedded in the jet since when the baroclinic front moves further downstream, the signal becomes weaker and is lost later.

From the data plotted in figure \ref{fig:Div1stream}(c), we can finally proceed to the Fourier space and analyse the horizontal wavelengths and frequencies depicted in figure \ref{fig:Div1stream}(d) \citep{yarom2014experimental}. , The wavenumber of the divergence along the streamline is on the $x$-axis, while the observed frequencies normalised by the Coriolis frequency $\omega_{M}/f$ are on the $y$-axis. The first feature that can be noticed is an asymmetry in the distribution of the peaks, showing a prominent signal on the right-hand side of the plot, i.e. for positive wavenumbers. The explanation for this is that the measured frequency is Doppler shifted, which means that it is the result of the intrinsic frequencies of the gravity waves summed to the one at which the flow propagates
\begin{equation}
 \omega_{M}=\omega_{i}+\vec{U}_{0} \cdot \vec{K}_{H}. \label{eq:DopplerMeasured}
\end{equation}

According to (\ref{eq:DopplerMeasured}), the asymmetry can be interpreted as the preferential direction of propagation in the jet flow direction. For waves propagating against the flow, one would instead observe the negative branch. It follows from the dispersion relation that the frequencies span the gravity wave range  $f+\vec{U}_{0} \cdot \vec{K}_{H}<\omega_{M}<N+\vec{U}_{0} \cdot \vec{K}_{H}$. The solid and dashed black curves mark these upper and lower limits in \ref{fig:Div1stream}(d). The horizontal wavenumber is assumed to vary in the interval -15 cm$^{-1}$ to 15 cm$^{-1}$, which is the range spanned by the measured waves. The Doppler shift calculated considering $U_{min}=0.2$ cm s$^{-1}$ and $U_{max}=0.5$ cm s$^{-1}$, which are the maximum and minimum velocity magnitudes measured along the jet, is added to $f$ and $N$. The two resulting curves delimit the region with the most significant amount of energy, indicating that the energy peaks are, indeed, related to gravity waves. \\

Although some IGW generation mechanisms relevant to the atmosphere can be excluded in the laboratory experiment, an exhaustive investigation of all the possible sources is challenging, as the full three-dimensional velocity field is not available. The horizontal divergence field is an appropriate choice for identifying inertia-gravity waves and study their characteristics, but it has the limitation that the field could still incorporate a non-negligible part of the balanced flow. A more precise separation of the balanced and unbalanced part of the flow becomes, therefore, of fundamental importance when studying spontaneous imbalance processes. Sophisticated methods involving a modal decomposition of the full field to separate balanced and unbalanced fields have been applied, for example, by \citet{hien2018spontaneous} for a linear system and by \citet{kafiabad2017rotating} and \citet{chouksey2018internal} for nonlinear systems. Unfortunately, these diagnostic tools require the full three-dimensional velocity field and are, therefore, often impossible to use for laboratory data or observations that generally do not give the full three-dimensional information. For this reason, with our experimental data, we can only investigate on whether some sources can be excluded and speculate on which sources are more likely to be responsible for the emission of IGWs. 

Among the possible sources of gravity waves in the rotating annulus, there are shear instabilities. To investigate such instability, the Richardson number $Ri=(N/\partial_{z}U)^2$ is obtained from the buoyancy frequency $N$, which is calculated by the vertical temperature profile measured with sensors, and the velocity shear ($\partial_{z}U$) calculated by PIV measurements at different fluid heights. The resulting values are $9.7<Ri<39$, a range that is much higher than $Ri_{\text{critical}}=0.25$ assuming horizontal homogeneity and a constant $N$. Although the measurement points do not cover the entire domain, and therefore some regions close to the walls might still represent an exception, shear instabilities can be ruled out for the investigated regions of the baroclinic wave (see also \citet{von2018instabilities} in this context). 

Another possible source of gravity waves in the experiment comes from the boundary instabilities at the inner cylindrical wall \citep{jacoby2011generation,randriamampianina2015inertia,von2018instabilities, hien2018spontaneous}. Although waves have been reported to be generated from the inner wall and then propagate towards the middle part of the gap, we speculate that such a mechanism is not of primary relevance for our experiment. Our speculation relies on the fact that the wave characteristics observed in our case differ from the ones observed by \citet{von2018instabilities} in a small tank. More specifically, \citet{von2018instabilities} gravity waves propagate retrograde with respect to the baroclinic flow while the opposite is true in our case. Furthermore, the waves in our experiment are continuously generated along the baroclinic jet and then propagate along with it rather than entering the jet from the inner wall. Finally, the numerical simulations by \citet{hien2018spontaneous} for a set-up of the differentially heated rotating annulus analogous to our experiment evinced that most of the waves in the core region of the baroclinic wave are spontaneously generated, even though their IGW fields differ from ours.

From the arguments exposed, we suppose that a substantial part of the IGWs is likely to be spontaneously emitted from the baroclinic jets. This thesis is also supported by \citet{Rodda2019}, where we showed a correlation between the size of the Rossby number and the region of IGW emission.
\section{Energy spectra}
\label{sec:spectra}
After discussing the characteristics of the gravity waves emitted from the baroclinic jet, we now move to the investigation of the energy distribution among the scales. The idea is to analyse the experimental data with the same methods used for atmospheric data and see if the overall spectra show the same global characteristics. The reference atmospheric spectra considered here are from \citet{nastrom1984kinetic}, who calculated wind and temperature spectra from measurements at the mid-latitudes taken from over 6900 commercial air flights. These spectra have been extensively cited in the literature and referred to as a reference for many successive studies aiming to explain the underlying physical phenomena. The \citet{nastrom1984kinetic} wavenumber spectra are one-dimensional; for this reason, we need to reduce our two-dimensional time series to a comparable dataset. 
As we discussed in the previous section, one of the main characteristics of the observed IGWs is their propagation along the jet, mostly advected by it. This peculiar behaviour allows us to apply Taylor's hypothesis \citep{taylor1938spectrum} if we consider the jet as a mean velocity field of constant velocity $U_{\text{max}}$. In this way, the transformation $E(f)=E(k)(2\pi)/U_{\text{max}}$ applies, where $E(f)$ is the frequency power spectrum and $E(k)$ the one-dimensional wavenumber spectrum and $f \approx U_{\text{max}}k/2\pi$ \citep{kumar2018applicability}.
Therefore, we can use the measured velocity time series to calculate the frequency spectra and then convert them into wavenumber spectra. The dataset is prepared by removing the data close to the boundary in the 2D plane (see figure \ref{fig:Div1stream}(a)), which are affected by larger uncertainties. This operation leaves us with a dataset consisting of 8132 time series. The power spectral density for each horizontal velocity component ($\hat{C_{u}}$ and $\hat{C_{v}}$) is calculated for every point in the 2D plane. Subsequently, a mean spectrum is derived from all the spectra so that local effects are smoothed out. Similar averaging procedures have also been applied by \citet{nastrom1984kinetic} and \citet{callies2014transition}.

\begin{figure}
  \centering
\includegraphics[width=0.7\linewidth]{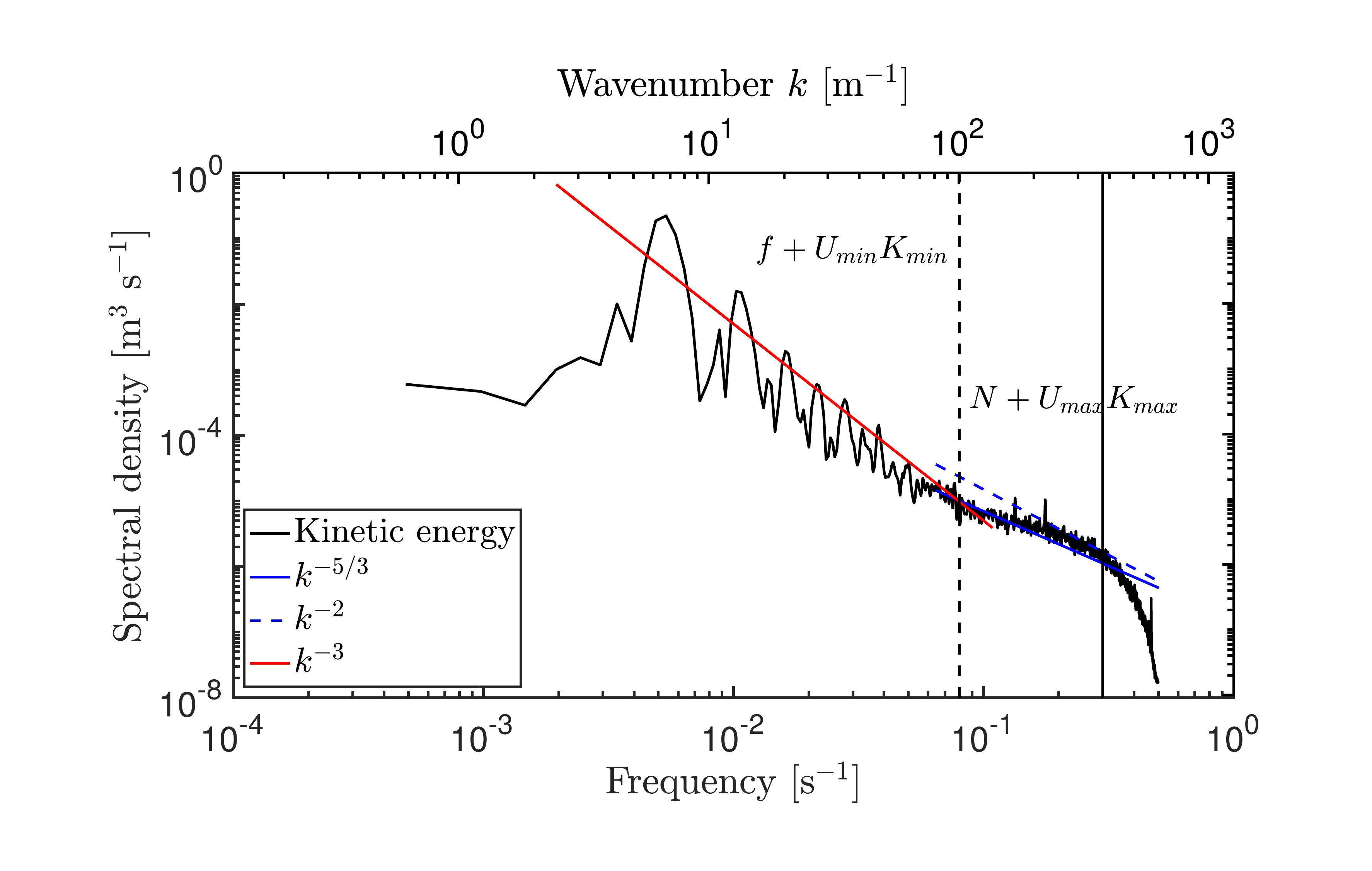}
   \caption[Mean kinetic energy spectrum.]{Mean kinetic energy spectrum. The red line displays the $k^{-3}$ power law, the blue line is $k^{-5/3}$, and the blue dashed line is $k^{-2}$. The vertical dashed and solid lines show the minimum and the maximum doppler shifted frequencies for inertia-gravity waves respectively. The minimum dashed line is calculated $\omega_{min}=f+U_{min}K_{min}$, where $f=0.02$ s$^{-1}$ and the doppler shift $U_{min}K_{min}=0.06$ s$^{-1}$. The maximum solid line is calculated $\omega_{max}=N+U_{max}K_{max}$, where $N=0.035$ s$^{-1}$ and the doppler shift $U_{max}K_{max}=0.24$ s$^{-1}$. Note that these lines are analogous to the curves plotted in figure \ref{fig:Div1stream}(d) for constant $K_{min}$ and $K_{max}$ measured.}
   \label{fig:spectra}
\end{figure}
The kinetic energy spectra have been calculated for different experimental runs where the lateral temperature difference is kept constant to $\Delta T=4.8$ K, but the rotation rate varies in the range $\Omega=0.5$ rpm to 1 rpm. The spectra for all the cases (not reported here for brevity) look qualitatively similar to the one generically chosen ($\Omega=0.5$ rpm) in figure \ref{fig:spectra}. A common problem that emerged from the analysis is that more extended measurements would be needed to resolve the low-frequency part of the spectra fully. Nevertheless, in all the measurements, there is a clear transition from a steeper to a shallower slope. In the following, we will discuss further results for the selected case that shows a steady baroclinic wave $m=6$. The analysis has been done for all the other dataset as well and, since the results are comparable, shall not be reported in detail here. \\

As previously discussed, the kinetic energy spectra calculated for our experiment (figure \ref{fig:spectra}) reveal a striking resemblance with typical atmospheric spectra \citep{nastrom1984kinetic}. Although the large scales are not entirely resolved, as it can be seen from the large fluctuations at low frequencies, an evident change of slope is visible at frequencies around $\omega=8 \times 10^{-2}$ s$^{-1}$. For the flatter part, two slopes are depicted: $-5/3$ and $-2$. The latter is the slope based upon the theory by \citet{garrett1979internal} for internal gravity waves in the ocean. We mention the Garret-Munk model because the internal gravity wave model for the atmosphere that we use is, in fact, an extension of the well established oceanic model. \citet{vanzandt1982universal} showed that for the atmospheric spectra, the slope should be changed from $-2$ to $-5/3$. Although the two slopes are not very far from one another, as it can be seen in figure \ref{fig:spectra}, the $-5/3$ fits our data almost perfectly. The better fit of the $-5/3$ slope suggests that, concerning the energy spectrum in the investigated regime, the experiment has more features in common with the atmosphere than with the ocean. The steepening of the spectrum for the highest frequencies, bringing forward that the measurements resolve the small scales up to dissipation. The dashed and solid grey vertical lines in figure \ref{fig:spectra} mark the range of frequencies of IGWs. The lower limit, indicated by the dashed line, is calculated $\omega_{min}=f+U_{min}K_{min}$, where $f=0.02$ s$^{-1}$ and the doppler shift $U_{min}K_{min}=0.06$ s$^{-1}$. The upper limit, indicated by the solid line, is calculated $\omega_{max}=N+U_{max}K_{max}$, where $N=0.035$ s$^{-1}$ and the doppler shift $U_{max}K_{max}=0.24$ s$^{-1}$. Note that these lines are analogous to the ones plotted in figure \ref{fig:Div1stream}(d), but this time the wave number is considered constant with minimum and maximum values ($K_{min}$ and $K_{max}$) evaluated from figure \ref{fig:Div1stream}(d). It can easily be seen that the flatter subrange of the spectrum lies in the IGWs frequency range. This already hints that gravity waves might substantially contribute to the energy at these scales.

\subsection{Spectra decomposition}
To understand better the energy distribution over the multiple scales present in the laboratory experiment, we use the spectral decomposition method developed by \citet{buhler2014wave} that allows separating the spectral part coming from the divergent and rotational part of the flow. A similar approached has been proposed by \citet{lindborg2015helmholtz}, and both methods have been applied to data collected in the atmosphere by aircraft (MOZAIC dataset). These two decomposition methods have been tested by \citet{bierdel2016accuracy} on identical datasets, and the results are almost indistinguishable. Based on this equivalence, we opted for the method developed by \citet{buhler2014wave} because it offers a further decomposition in wave and vortex components of the flow, based on the assumption that gravity waves are in a nearly linear regime.

The wave vortex decomposition method by \cite{buhler2014wave}, consisting of a two-step decomposition, is briefly described first and then applied to our experimental data.
The first step is a Helmholtz decomposition performed over the horizontal velocity power spectra ($\hat{C}^{u}$ and $\hat{C}^{v}$) to separate the rotational $K^{\psi}(k)$ and the divergent component $K^{\phi}(k)$ in the following form
\begin{subequations}
\begin{align}
K^{\psi}(k)&=\frac{\hat{C}^{v}(k)}{2}+\frac{1}{2k} \int_{k}^{\infty}\left( \hat{C}^{v}(\overline{k})-  \hat{C}^{u}(\overline{k}) \right ) d\overline{k},\label{eq:kpsi} \\
K^{\phi}(k)&=\frac{\hat{C}^{u}(k)}{2}-\frac{1}{2k} \int_{k}^{\infty}\left( \hat{C}^{v}(\overline{k})-  \hat{C}^{u}(\overline{k}) \right ) d\overline{k}. \label{eq:kphi}
\end{align}
\end{subequations}

Since we have a discrete data set, the integrals are calculated numerically via the trapezoidal method implemented in the \textsc{Matlab} function \textbf{trapz}, where the integration area is broken down into trapezoids spaced by the distance between each sampled point.
\begin{figure}
  \centering
\includegraphics[width=0.73\linewidth]{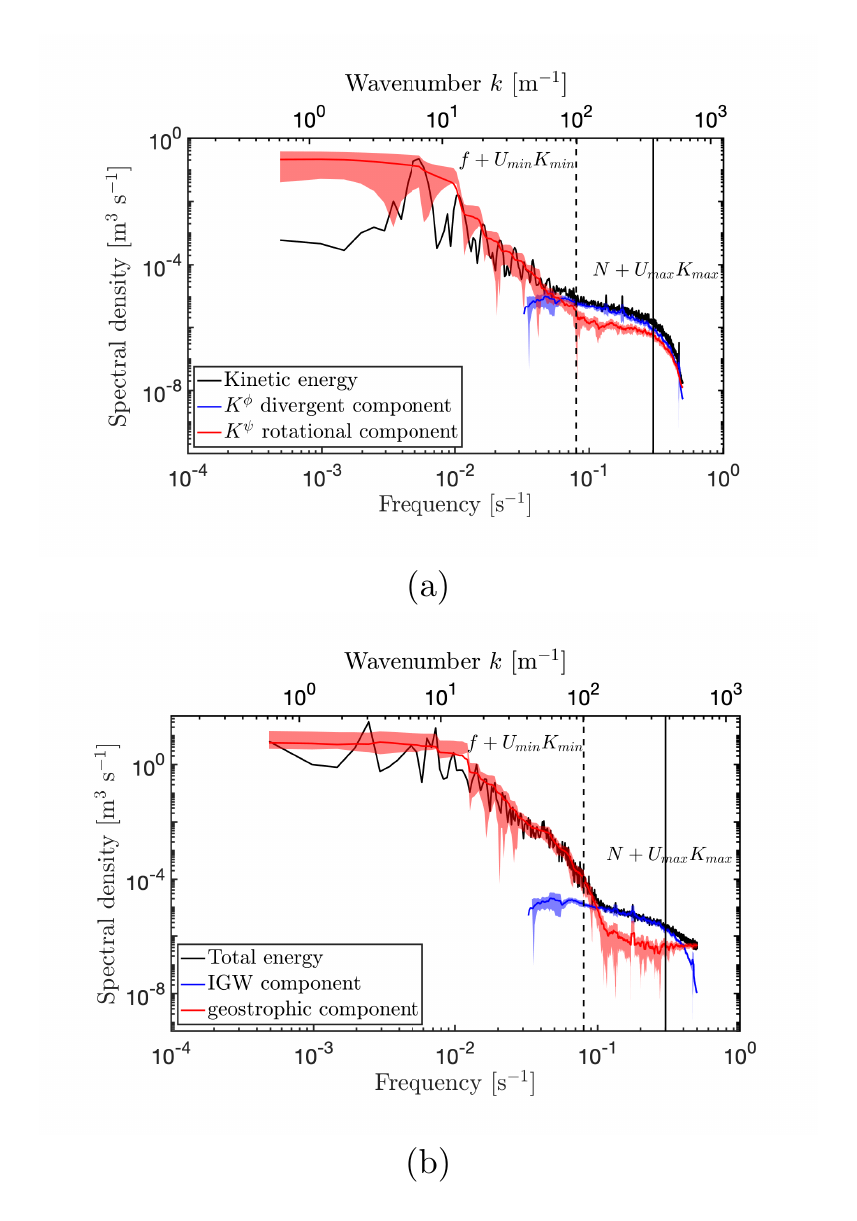}
   \caption[Energy spectra decomposition.]{Energy spectra decomposition. (a) Kelvin-Helmholtz decomposition; the black curve shows the spectra of the kinetic energy, the red curve is the rotational component (\ref{eq:kpsi}), and the blue curve is the divergent component (\ref{eq:kphi}). (b) gravity waves-residual geostrophic component decomposition; the black curve shows the total energy spectrum calculated by adding the potential energy (\ref{eq:potentialSpectra}) to the kinetic energy spectrum, the blue curve is the wave component $E_{W}(k)$ calculated in (\ref{eq:wavecomponent}), and the red curve is the residual geostrophic component calculated by subtracting $E_{W}(k)$ from the total energy. The dashed and solid line are calculated as in figure \ref{fig:spectra}. In both plots, the red and blue curves show the data smoothed with a moving average calculated over 20 data points, while the colored shadows represent $2\sigma$ variance. }
   \label{fig:spectra-decomposition}
\end{figure}

Figure \ref{fig:spectra-decomposition}(a) presents the Helmholtz decomposition for the energy spectra, obtained from (\ref{eq:kpsi},\ref{eq:kphi}). The kinetic energy is plotted in black, the rotational component $K^{\psi}$ (\ref{eq:kpsi}) in red, and the divergent component $K^{\phi}$ (\ref{eq:kphi}) in blue. The blue and the red curves intersect at a frequency equal to $4 \times 10^{-2}$ s$^{-1}$, which corresponds to the frequency where the slope flattens in figure \ref{fig:spectra}(c). Another feature visible in figure \ref{fig:spectra-decomposition}(a) and later in figure \ref{fig:spectra-decomposition}(b) is that the divergent energy spectrum, attributed to the subsequently calculated wave energy spectrum, is truncated at small frequencies. This truncation occurs because the decomposition method is based on the assumption that the velocities are isotropic. As \citet{buhler2017anisotropic} recently evinced, if the data are anisotropic, the divergent energy spectrum becomes negative at small wavenumbers and this results in the gap of data, similar to what is observed in the plots presented here. To resolve this problem, \citet{buhler2017anisotropic} developed a method that, under certain conditions, can be applied to anisotropic data. This anisotropic method involves the further calculation of the cross-spectrum $\hat{C}^{uv}$, and it can be applied under the assumptions that the imaginary part $\Im({\hat{C}^{uv}})$ is negligible and the real part $\Re({\hat{C}^{uv}})$ is sign-definite. However, both conditions are not fulfilled for our data set, and therefore the anisotropic method cannot be used. Even so, the isotropic separation method gives results that fully cover the flatter part of the spectrum (on which our analysis is mostly focussed) and additionally reaches deep into the frequency part where the rotational component has a steep slope. Hence, the region where the rotational and the divergent part intersect is well resolved. 

The rotational component (red line in figure \ref{fig:spectra-decomposition}a) is the one contributing the most to the energy for small frequencies, while the divergent component (blue line) becomes more important at frequencies in the range of the gravity waves. If we translate frequencies into wavenumbers (visible in the top $x-$axis in figure \ref{fig:spectra-decomposition}(a) and (b)), this plot clearly shows that the divergent component matters at small scales. The experimental spectra (figure \ref{fig:spectra-decomposition}a) differ from the ones reported in \citet{callies2014transition} because our spectra are well separate even in the region with the $-5/3$ slope. In summary, our experiment directly supports the idea that the divergent part of the atmospheric spectrum is due to gravity waves and not to stratified turbulence. The latter would be the case if the rotational and divergent component would be in the same order for the IGW frequency range \citep{lindborg2007horizontal}.\\

The second step of the \citet{buhler2014wave} method consists in decomposing the total energy spectrum into the geostrophic and inertia-gravity wave components. With respect to the Helmholtz decomposition, this further step accounts for inertia-gravity waves made of both a divergence and a rotational component, and it is based upon the linear dynamics of inertia-gravity waves (we shall discuss the validity of this assumption for our data in the next section). For calculating the total energy spectrum, we need to consider the potential energy in addition to the already calculated kinetic energy. 
From the temperature measurements collected via temperature sensors placed in the middle of the tank, we could determine the corresponding potential energy spectrum, which reads

\begin{equation}
\hat{C}^{b}(k)=|\hat{b}(k)|^{2}/N^{2}, \label{eq:potentialSpectra}
\end{equation}
where $b$ is the buoyancy and $N$ the buoyancy averaged over space and time.

By further assuming that the field is composed of plane waves, from which vertical homogeneity follows, we can use the dispersion relation of gravity waves (\ref{eq:IGWdisp}) to derive the equipartition relation for the spectra as
\begin{equation}
\hat{C}^{b}(k)+2K^{\psi}(k)=2K^{\phi}(k)+\hat{C}^{w}_{W}(k) \approx E_{W}(k), \label{eq:wavecomponent}
\end{equation}
where $\hat{C}^{w}_{W}(k)$ is the vertical kinetic energy, which we will assume negligible. With this additional assumption, the hydrostatic internal wave energy spectrum is $E_{W}(k)\approx 2K^{\phi}(k)$. 

The total energy decomposition in figure \ref{fig:spectra-decomposition}(b) further confirms that the small scales are dominated by nearly linear gravity waves, whose energy is comparable to the total energy spectrum in the $-5/3$ range. The comparison of figure \ref{fig:spectra-decomposition}(a) and \ref{fig:spectra-decomposition}(b) evinces similar energy parts for the divergence and the inertia-gravity wave component, also indicating that here the horizontal component is the most energetic part of the gravity waves.\\

In both decompositions the rotational (see red curve in figure \ref{fig:spectra-decomposition}(a)) and geostrophic (see red curve in figure \ref{fig:spectra-decomposition}(b)) component keep having a steep slope at the mesoscales, but do eventually flattens towards a $-5/3$ slope parallel to the divergent and gravity wave spectral component  (see red curve in figure \ref{fig:spectra-decomposition}(a) and (b) respectively). An analogous behaviour has been reported by \citet{callies2014transition} and the authors described this flattening as an artifact and attributed this to interpolation procedure and truncation errors. On the other hand, some numerical studies reported analogous features with the geostrophic spectrum approaching a $-5/3$ slope for very small scales and associated this behaviour to a transition towards an isotropic turbulence regime \citep{kafiabad2016balance, bartello2010}.
\\
\subsection{Weakly nonlinear regime approximation}
\label{sec:linear}
We have seen in the previous section that the kinetic energy spectrum consists of a divergent and a rotational part $E_{\text{kin}}(K)=K^{\phi}+K^{\psi}$. The ratio between the two components, defined as
\begin{equation}
R=\frac{K^{\phi}}{K^{\psi}}, \label{eq:R}
\end{equation}
gives another essential piece of information about the dynamics at the mesoscale. Indeed, realistic frequency distributions of linear gravity waves are expected to give  $R>1$ in the mesoscale. For $R<1$ linear atmospheric IGWs are ruled out as primary contributors to the kinetic energy since the dynamics is dominated by vortical flows \citep{li2018weakly}. It is also evident that for quasi-geostrophic dynamics $K^{\phi}$ should be very small compared with $K^{\psi}$, therefore in regions where $R<<1$ balanced regimes are dominating. For $R \simeq 1$ stratified turbulence is a possible candidate.

\begin{figure}[t]
  \centering
\includegraphics[width=0.7\linewidth]{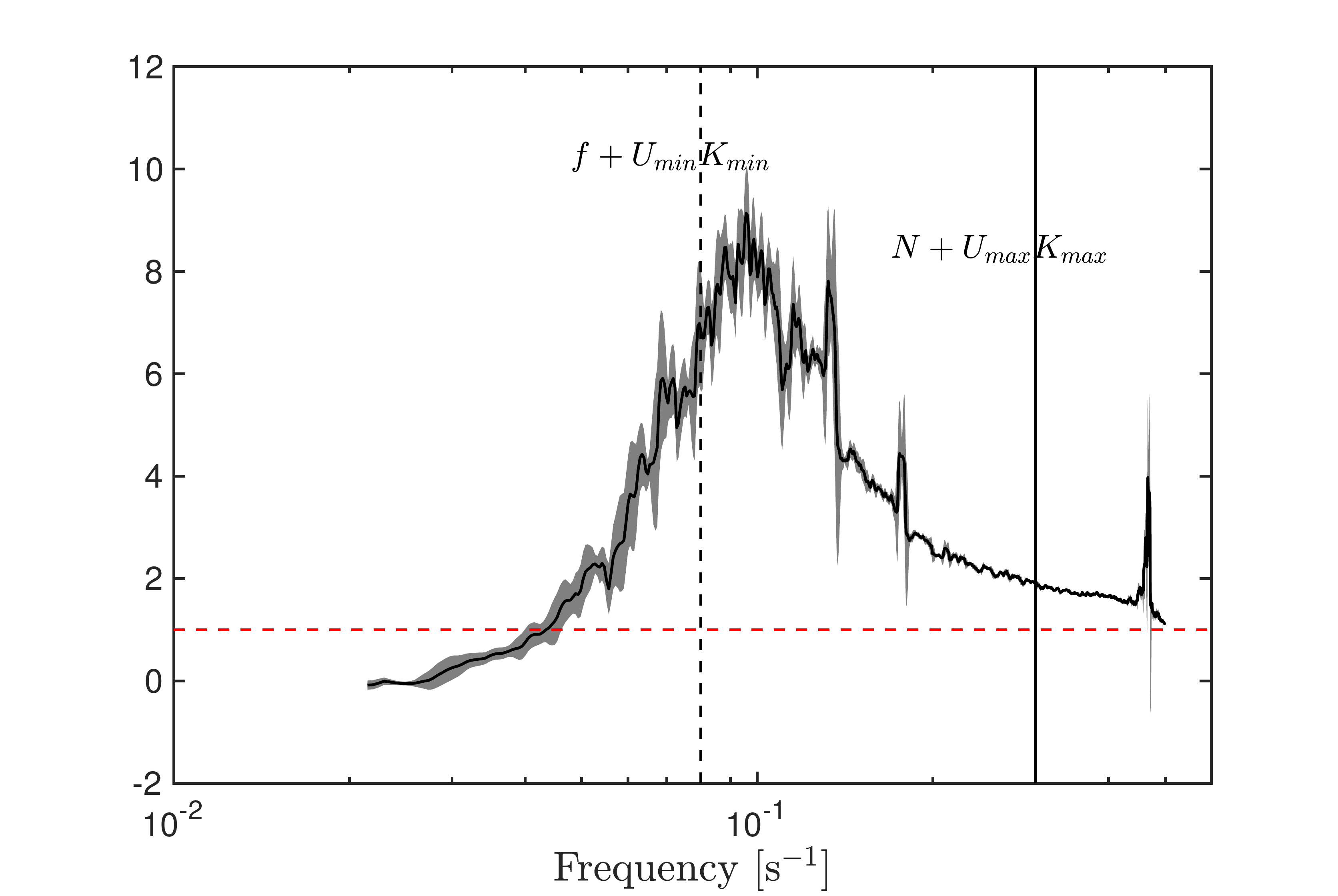} 
   \caption[Ratio between divergent and rotational energy spectra]{Ratio between divergent and rotational energy spectra calculated from (\ref{eq:R}). The black solid line shows $R$ smoothed with a moving average over 10 data points, while the gray shadow shows the variance $2\sigma$. The red horizontal dashed line indicates $R=1$. The vertical dashed and solid black lines indicates $f+U_{\text{min}K_{\text{min}}}$ and $N+U_{\text{max}K_{\text{max}}}$ to facilitate the comparison of the frequency ranges with figures \ref{fig:spectra-decomposition} (a) and (b).}
   \label{fig:R}
\end{figure}
\begin{figure}[t]
  \centering
\includegraphics[width=0.7\linewidth]{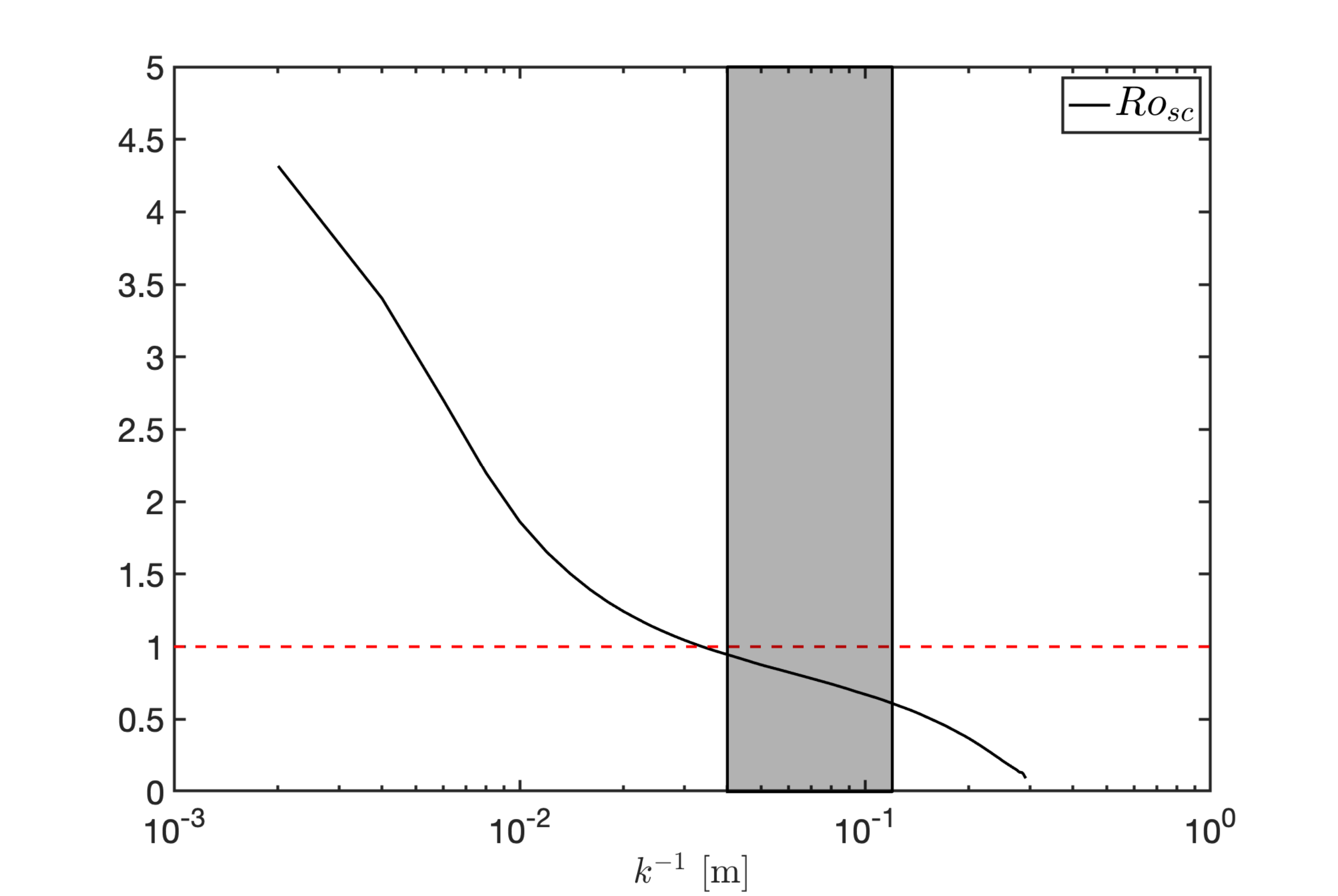}
   \caption[Scale dependent Rossby number]{Scale dependent Rossby number calculated with (\ref{eq:Ro-scale}). The black solid line shows $Ro_{\text{sc}}$. The horizontal dashed red line pinpoints $Ro=1$. The gray shaded area marks the scales between the IGW wavelengths and the Rossby deformation radius $L_{D}$.}
   \label{fig:RoSpectra}
\end{figure}
Figure \ref{fig:R} shows the $R$ variations over the frequency range; the vertical dashed and solid lines indicating the lower and upper limits for the IGW frequency range are plotted as a reference. The smallest frequency considered is $\omega_{\text{min}}=1 \times 10^{-2}$ s$^{-1}$ since for $\omega<\omega_{\text{min}}$ the divergent component has imaginary values that cannot be considered (see also figure \ref{fig:spectra-decomposition}(a)). For $1 \times 10^{-2}$ s$^{-1}$ $<\omega<4 \times 10^{-2}$ s$^{-1}$ the ratio $R$ lies below the red dashed line, which indicates $R=1$. This part is dominated by the large-scale balanced flow, which is in the quasi-geostrophic regime. For $\omega>4 \times 10^{-2}$ s$^{-1}$ the divergent component grows much larger and, therefore, we find $R>1$. Since the slope in figure \ref{fig:spectra} becomes shallower for the same frequency range, the large values for $R$ confirm that gravity waves are, indeed, the main dynamical process underlying the energy at the small-scales. For larger frequencies, starting at $\omega \approx 2 \times 10^{-1}$ s$^{-1}$, the ratio $R$ decreases and tends towards $R=1$. At these smaller scales, we have previously seen that also $K^{\psi}$ enters the shallow part and approaches the $-5/3$ slope as well. If this region corresponds to a turbulent regime, the energy should be equally distributed among  the geostrophic and the gravity wave modes from which it follows that the geostrophic and ageostrophic spectra should have the same slope \citep{bartello2010}. A possible interpretation of our finding is that the balanced flow transfers first energy to the imbalanced flow in the form of gravity wave emission. Such waves have smaller scales but are still within the order of magnitude of the Rossby deformation radius and the size of the baroclinic vortices. At even smaller scales there is a forward cascade from the gravity wave scale to even smaller scales that eventually lead to dissipation. In a recent paper by \citet{kafiabad2019diffusion}, the relevance of IGW diffusion by turbulence is investigated, and evidence that this mechanism produces a $-5/3$ slope of the spectra is reported. The proposed mechanism is consistent with our experimental data.\\

Another critical quantity to look at for establishing whether the energy spectra are generated by weakly or strongly nonlinear dynamics is the Rossby number. More in detail, the weak nonlinearity condition is verified if the Rossby number is small. Following \citet{li2018weakly}, we consider a scale-dependent Rossby number defined as
\begin{equation}
Ro_{\text{sc}_{i}}=\frac{\sqrt{<\delta \bold{U} \cdot \delta \bold{U}>}}{f dx_{i}} \label{eq:Ro-scale}
\end{equation}
where the brackets $<>$ indicate a spatial average, $\delta \bold{U}=\bold{U}_{x_1}-\bold{U}_{x_2}$ is the velocity difference between two velocities measured at the points $x_1$ and $x_2$, $dx=x_1-x_2$ is the distance between such points, and $f$ is the Coriolis parameter. To calculate $Ro_{\text{sc}_{i}}$, we used the radial velocity component and calculated the differences between points along the radial direction. The same operation is started with two adjacent velocities and then repeated increasing stepwise the distance between them until the two points are at the opposite extremes of the domain. This procedure allows us to investigate a range $0.2$ cm$<dx<$ $30$ cm, i.e. from the spatial resolution of our data almost to the entire gap width (some data at the boundaries have been removed because of the significant noise level). The calculation is repeated for the entire time series, and then the mean value is taken.

The resulting $Ro_{\text{sc}}$ is plotted in figure \ref{fig:RoSpectra}, where the red horizontal dashed line marks the value $Ro=1$. The experimental $Ro_{\text{sc}}$ shows strong similarities with scale-dependent Rossby number for stratospheric data (see figure 7 in \cite{li2018weakly}). For the smallest scales, $Ro>1$ implying that the dynamics is strongly nonlinear. At scales larger than 3 cm, the Rossby number transitions to values smaller than one. The region by the grey shading marks scales between the observed IGW wavelengths and the Rossby deformation radius, i.e. the typical dimension of baroclinic vortices. It follows that at scales for which the Rossby number is smaller than one and $R>1$, linear gravity waves are likely to be responsible for the $k^{-5/3}$ spectrum. These findings are in agreement with \cite{li2018weakly}, who concluded that part of the mesoscale spectrum in the lower stratosphere might result from linear gravity waves.

Finally, the marked transition to a nonlinear regime for scales smaller than 3 cm is in agreement with our hypothesis that there is a transition from nearly linear gravity waves towards a turbulent dynamics at the smaller scales.
%

\section{Conclusions}
\label{sec:conclusions}
We have presented in this paper an experimental investigation of inertia-gravity wave (IGW) emission from the balanced baroclinic flow in the differentially heated rotating annulus. IGWs are identified in the horizontal divergence field and appear to be emitted from the jet front and subsequently to propagate along with it, mostly advected. The waves in the experiment show several similarities with the waves observed in the atmosphere \citep{suzuki2013evidence, dornbrack2018gravity}, with the wave crests oriented perpendicularly to the jet as a consequence of the wave-capture process they undergo when propagating through regions of the flow with large shear deformation \citep{plougonven2005gravity}.

One open issue is understanding the contribution of IGWs in the mid-latitudes to the total energy at the mesoscales. This problem has been debated for more than thirty years, but the analysis of atmospheric data still shows controversial results.
Although the experiment has several simplifications compared to the real atmosphere, the one-dimensional energy spectra obtained from a series of experimental runs in a regular baroclinic wave regime show striking similarities with the atmospheric spectra. The characteristic $-3$ slope for the large-scale balanced flow turning into a $-5/3$ slope for smaller scales is well visible in the kinetic energy spectra. A Helmholtz decomposition is applied to such spectra and reveals that most of the large-scale energy is attributed to the balanced rotational flow, while at the smaller scales, in the sub-range with slope $-5/3$, the divergent component of the spectrum becomes the most energetic. The ratio between the divergent and the rotational component $R$ has a peak at $R=8$ in the IGW range and then lowers approaching 1 for smaller scales, corresponding to the spectral range where the rotational part approaches a slope of $-5/3$. 

A possible interpretation of these findings is that the balanced flow undergoes a loss of balance and is spontaneously emitting inertia-gravity waves to the mesoscale. This process is signalled by a slope change in the energy spectrum, where part of the rotational mode energy is transferred to the divergent modes. Hence, the dynamics at the mesoscale is dominated by inertia-gravity waves. In the $-5/3$ region, the energy is cascading forward, being transferred from larger to smaller scales until it dissipates by diffusive processes. Therefore, a further change in the dynamics might occur corresponding to an observed flattening of the vortical component. 

Our experimental findings also seem to confirm the results obtained through numerical simulations of an idealised baroclinic wave life cycle by \citet{waite2009mesoscale}. The numerical simulations reveal that spontaneously emitted gravity waves are responsible for the shallower part of the stratospheric spectra (although the same is not observed for the tropospheric spectra). 
One reason for the alignment of our results with the numerical simulations related to the stratosphere is that, since the lower stratosphere is calmer than the upper troposphere, the stratospheric dynamics might be closer to the regimes we considered in the experiments. Nevertheless, to what extent the processes observed in the experiment apply to the atmosphere is an open issue. 
One fundamental advantage of doing baroclinic life cycle studies by laboratory experiment is that all the scales up to dissipation are resolved. In the numerical models, instead, the scale smaller than the transition need to be parameterized, and the numerical resolution could significantly influence the mesoscale flow.  

In conclusion, we have shown that the differentially heated rotating annulus with a small aspect ratio is not only an analogue to the \emph{large-scale} atmospheric flow, but can also be a useful tool to study multiple-scale processes in atmosphere-like flows and validate numerical simulations and field data. 

%
\section{Acknowledgments}
This work was supported by the Spontaneous Imbalance project (HA 2932/8-1 and HA 2932/8-2) that is part of the research group Multiscale Dynamics of Gravity Waves funded by DFG (FOR1898). The authors thank Ludwig Stapelfeld, Robin St\"obel, and Florian Pr\"{u}fer for technical support. Moreover, we thank the Spontaneous Imbalance group of MS-GWaves, (Ulrich Achatz, Ion Dan Borcia, Steffen Hien, Lena Schoon, Christoph Z\"{u}licke) for support and fruitful discussions.

%

 \bibliographystyle{ametsoc2014}
 \bibliography{mybibliography}


\end{document}